# Canada Protocol: an ethical checklist for the use of Artificial Intelligence in Suicide Prevention and Mental Health

**Carl-Maria Mörch**, M.Psy., Ph.D. Student[1]

Contribution : Lead researcher, study design, writing, coordinated the validation

**Abhishek Gupta**, B.A.[2]

Contribution : Co-author, co-designed the checklist, participated to the selection of the checklist's items.

**Brian L. Mishara**, Ph.D.[3]

Contribution : Scientific supervision, complete revisions

## Abstract

Introduction: To improve current public health strategies in suicide prevention and mental health, governments, researchers and private companies increasingly use information and communication technologies, and more specifically Artificial Intelligence and Big Data. These technologies are promising but raise ethical challenges rarely covered by current legal systems. It is essential to better identify, and prevent potential ethical risks. Objectives: The Canada Protocol - MHSP is a tool to guide and support professionals, users, and researchers using AI in mental health and suicide prevention. Methods: A checklist was constructed based upon ten international reports on AI and ethics and two guides on mental health and new technologies. 329 recommendations were identified, of which 43 were considered as applicable to Mental Health and AI. The checklist was validated, using a two round Delphi Consultation. Results: 16 experts participated in the first round of the Delphi Consultation and 8 participated in the second round. Of the original 43 items, 38 were retained. They concern five categories: "Description of the Autonomous Intelligent System" (n=8), "Privacy and Transparency" (n=8), "Security" (n=6), "Health-Related Risks" (n=8), "Biases" (n=8). The checklist was considered relevant by most users, and could need versions tailored to each category of target users.

Funding: This article has not been funded.

Conflict of Interests: The authors declare no conflict of interests.

Ethics: In agreement with the Université du Québec à Montréal's ethics board rules, this research did not include any sensitive or identifiable information on individuals.

---

[1] Centre for Research and Intervention on Suicide, Ethical Issues and End of Life Practices (CRISE)

Psychology Department, Université du Québec à Montréal
c.p. 8888, Succ. Centre-Ville, Montréal, Québec, H3C 3P8, Canada
cmmorch@gmail.com

[2] Montreal AI Ethics Institute, Canada

Microsoft, Montreal, Canada

[3] Centre for Research and Intervention on Suicide, Ethical Issues and End of Life Practices (CRISE)

Psychology Department, Université du Québec à Montréal

**Table 1** : The Canada Protocol – the ethical checklist

| **DESCRIPTION** | |
|---|---|
| **Objectives** | Describe your project's objectives and/or rationale and describe the role and functioning of your Autonomous Intelligent System |
| **Technology** | Name and describe the technologies and techniques used (e.g. supervised or unsupervised learning, machine learning, random forest, decision tree...). You can refer to the report of the AI Initiative incubated at Harvard http://ai-initiative.org/wp-content/uploads/2017/08/Making-the-AI-Revolution-work-for-everyone.-Report-to-OECD.-MARCH-2017.pdf. Mention the names of any technological intermediary or supplier allowing you to use the technology (e.g. technical provider, cloud provider) |
| **Funding & conflict of interest** | Indicate all sources of funding for your project (public and private) and who might have an interest (e.g. financial, political) in your Autonomous Intelligent System |
| **Credentials** | If you have noted that you or someone in your team has an expertise in relation to the Autonomous Intelligent System (e.g. in a document, a webpage, an interview), clearly indicate the name of the professional, their technical, academic or medical credentials, and their training (e.g. "Professor Smith, PhD in computer systems engineering from Harvard University. Specialist in the Online Detection of Depression") |
| **Target population** | Describe your target population and its size, or identify its subgroups and their sizes. Describe if and how the target population (and, or its subgroups) assisted in the design of your Autonomous Intelligent System. |
| **Evidence** | If you made claims about your Autonomous Intelligent System's efficacy, performance, or benefits, please justify them and provide the evidence underlying them. If you have mentioned or used scientific papers, please cite your sources |
| **Testing** | If you have run your Autonomous Intelligent System under adversarial examples or worst-case scenarios, describe the type of tests used and their outcomes |
| **Complaints** | Describe the process whereby users can formally complain or express their concerns about your Autonomous Intelligent System |
| **PRIVACY & TRANSPARENCY** | |
| **Responsibility** | Describe who will be legally accountable for your Autonomous Intelligent System's actions or decisions |
| **Data collection** | Describe what data have been collected and used (for the training, evaluation and operational phases), where they are stored, who collected the data, who will have access to the data, and what safeguards are in place to ensure secure storage |
| **Accessibility** | In all the documents or texts, confirm that you have used a language adapted to target users and, when relevant, accommodated special needs some users may have. |
| **Informed consent** | State whether you have obtained informed consent and, if so, how, when, and from whom. Describe its nature (formal, implied, renewable, dynamic) and include the exact wording on the consent form. Note whether you have received ethical approval from an institution (eg: hospital, university) for your consent forms |
| **Consent withdrawal** | State whether you have specified the duration of the consent and whether you have implemented consent withdrawal mechanisms (e.g. opt-out clause, unsubscribe option). Specify what happens if an user wants to stop using the AIS or delete his or her information |
| **Access to the data** | Access to the data: State if an individual can access any data related to him or her and obtain the data in a clear and structured export document. If this is not possible, explain why |
| **Right to be forgotten** | Describe whether an individual can retrieve and erase all of his or her information, and if so, how. Describe the mechanism |
| **Minors** | Note whether information concerning minors is used for the Autonomous Intelligent System. If it is, and it is intentionally collected, please indicate whether parental consent is required. If it is, and it is unintentionally collected, please describe what can be done to remove this information |
| **SECURITY** | |
| **Embedded recording mechanism** | If you have used a technology to monitor and record all your Autonomous Intelligent System's decisions and actions, detail how and in what circumstances these records could be made available to authorities, external observers or auditors |
| **Third-parties** | Indicate who has access to the data (individuals and organizations), and whether identifying information about participants is included in accessible data |
| **Data protection** | Detail all the measures taken to protect any sensitive and personal information |
| **Audit trails** | Explain who has access to the data and when |
| **Autonomy** | Explain if your system has the autonomy to take actions or make decisions on its own. If yes, detail the degree of autonomy of your Autonomous Intelligent System (e.g. partial or complete) |
| **Moderation** | Explain if your Autonomous Intelligent System requires human intervention or moderation. If yes, describe who will have access to your Autonomous Intelligent System, and what will the guides regulating their intervention be |
| **HEALTH-RELATED RISKS** | |
| **Type of care** | Is your Autonomous Intelligent System helping its owners to provide the target population with the optimal treatment or treatment as usual? Indicate the criteria (and their sources) for optimal treatment or treatment as usual |
| **Crisis & contigency planning** | List the criteria for evaluating the risk exposure of your Autonomous Intelligent System. Describe your plan in case of emergency, disaster, or suicidal crisis (the intervention protocol). If possible, specify what type of behaviors and environments are considered as being at risk and explain the rationale in a simple way |
| **Non-maleficience** | Explain whether your Autonomous Intelligent System could harm, incommode, or embarass a user and, if so, how. Explain how you avoid or minimize this risk |
| **Misuse** | Describe potential misuses of your Autonomous Intelligent System (e.g. describe a possible negative scenario to indicate what could potentially happen to a user) and describe your mitigation strategies |
| **Emotions detection** | If your Autonomous Intelligent System detects user's emotions, state how, and for what purpose. Explain whether the user is informed and if so, how |
| **Emotions control** | If your Autonomous Intelligent System can provoke emotions, describe how users are informed of this possibility, the emotions that may be provoked, their intensity, and possible impact on users |
| **Relationship** | Is the user aware that he or she is interacting with a machine? Describe whether your Autonomous Intelligent System can create a relationship with users, and if so, how. Describe how the relationship might affect a user |
| **Public awareness** | Describe the impact on users and potential users of public dissemination of information about your Autonomous Intelligent System and the process of its development |
| **BIASES** | |
| **Ethics** | If you have requested an expertise on ethics during the design of your Autonomous Intelligent System, detail the parties involved and their contributions |
| **Exclusion & discrimination** | Explain if there are risks of exclusion or discrimination related to your Autonomous Intelligent System (e.g. based on gender, race, age, religion, politics, health, sexual orientation, etc.) |
| **Stigmatization** | Describe how you avoided using languages, images, and other content that could stigmatize users (e.g., reference to guidelines on safe media reporting and public messaging about suicide and and mental illness) |
| **Detection** | If applicable, explain any potential detection errors that might be made by your Autonomous Intelligent System (e.g. false positives, false negatives) and estimate their extent (e.g. precision, recall). Describe any potential adverse consequences for users. If applicable, describe any incidental finding made by your Autonomous Intelligent System |
| **Data handling** | If applicable, describe the nature and purpose of any data manipulation (e.g. cleaning, transformation) and by whom they were performed. Describe what will be done with the metadata |
| **Data selection** | Describe where the data came from, how you accessed them (e.g. through an API) and if you think there might be a selection or sampling bias (e.g. the data comes from an API or a spectrum bias) |
| **Data transformation** | If applicable, describe the nature and purpose of any statistical transformations applied to your data. Describe any potential bias or risk related to the data transformation (e.g. ecological fallacy, confounding factors) |
| **Other** | If you have identified other potential methodological or scientific biases, describe them and their potential ethical consequences (e.g.1. an excessively long consent form could affect the informed consent; e.g.2. the presence of a floor effect in the measurements could constrain an Autonomous Intelligent System's ability to detect a behavior) |

# Introduction

Mental health care is deeply transformed by Information and Communication Technologies (ICT) and in particular Artificial Intelligence. It brings new exciting promises, from better diagnosing clients (Masri & Jani, 2012), to improving clinical decision making (Bennett & Doub, 2016; Bennett & Hauser, 2013).

This has become a major field of interest for multiple countries. For example, the Canadian government has recently promoted the concept of E-Mental Health (Mahajan et al., 2014). It is defined as a way to help "people living with mental health problem or illness feel to more knowledgeable and better express their needs" (Mahajan et al., 2014; MHCC, 2017). One specific domain benefits from this technological trend: suicide prevention (MHCC, 2018).

Mental Health and Suicide are indeed two interlinked themes. A 2014 World Health Organization Report considers suicide prevention to be the priority condition which is the focus of their Global Strategy for Mental Health (WHO, 2017). It also identifies several objectives: improving the way suicide rates are monitored, suicidal individuals are detected, and psychological help is delivered (Varathan & Talib, 2014; WHO, 2014). In order to meet these ambitious goals, governments (e.g. Canada), researchers and private companies (e.g. Facebook) are increasingly using Artificial Intelligence (Luxton, 2015; Mörch et al., 2018).

The term "Artificial Intelligence" designates a very wide field of study, which may elicit both ethical concerns and hopes for more effectively prevention suicides and treating mental health problems (CNIL, 2017). It is defined as the domain interested in the study and design of intelligent machines (McCarthy, 1998). Its goal is to "build machines that are capable of performing tasks that we define as requiring intelligence, such as reasoning, learning, planning, problem-solving, and perception" (Luxton, 2015, p. 15). The term has become popular, but is difficult to accurately define. A recent international report suggested it may be better to use the term: Autonomous Intelligent System (AIS) (IEEE, 2017). In the context of mental health care, some researchers have used the term, Intelligence Care Providers (ICP) (Luxton, 2015). There are also many types of AI varying in methods and terminology. The most widespread type of AI is its sub-branch: Machine Learning (Demiaux & Si Abdallah, 2017).

Generally speaking, the whole field is commonly associated with "Big Data" (BD). This term is defined as the constant increase of generated data and the development of techniques to analyze them (Kitchin, 2014). These two domains are often associated and raise numerous ethical questions. Many studies, books, and reports have been published on ethical issues concerning AI (Brundage et al., 2018; Demiaux & Si Abdallah, 2017; IEEE, 2017). It is important to note that almost all these ethical dilemmas are new and go beyond the scope of the present legal systems (Villani et al., 2017).

Suicide prevention is no exception: the use of AI poses many new ethical, technical, and scientific challenges. There is however a paradox. Recently published studies that use BD or AI in suicide prevention rarely mention encountering ethical issues or challenges (Mörch et al., 2018). This relative absence of ethics in such a scrutinized field appears surprising. Firstly, there is an identified challenge: current guidelines and ethics codes do not always cover the ethical issues that are pertinent to health care and Artificial Intelligence. Secondly, today computer engineers and experts from around the world now talk of the concept of "moral overload" (IEEE, 2017). This expression means that AI developers and users frequently feel overwhelmed by the magnitude of the ethical risks, and do not feel competent, nor trained to address those risks.

This article proposes a simple tool for people who want to use AI in mental health and suicide prevention: a checklist to identify and anticipate ethical issues, called the Canada Protocol - MHSP. This is the first validated instrument of its kind in AI Ethics. This tool, created by the authors, was designed to cover the challenges identified in the field of suicide prevention and mental health, in ethics and AI and ICT in health care. This paper presents the protocol and its validation process by an international committee of 16 professionals and experts, using a two-round Delphi Consultation.

The purpose of the creation of this checklist is to contribute to ethical education and to the improvement of practices when AI and BD are used in mental health care and suicide prevention. This checklist was constructed for use by AI developers, researchers, and decision makers willing to use AI in suicide prevention, and professionals and practitioners considering using AI.

Methodology

a. Checklist

The Canada Protocol - MHSP is a checklist. Checklists have long been used in mental health care, to elaborate list of potential symptoms in order to orient treatment (e.g. the Symptom 90 Checklist Revised (Derogatis, 1979; Derogatis & Unger, 2010)), or to better detect specific disorders (e.g. the Post-Traumatic Disorder Checklist) (Weathers, Litz, Herman, Huska, & Keane, 1993). Checklists are generally used to help clinicians, but they can also be used to help researchers assess the quality of studies (e.g. the National Institute for Clinical Excellence's checklist - NiCe) (Excellence, 2009). An important example is the *Preferred Reporting Items for Systematic Reviews* (PRISMA), which was designed by the RAND Foundation to improve the quality of systematic reviews (Moher, Liberati, Tetzlaff, & Altman, 2009). The PRISMA checklist is comprised of 27 items and a flow diagram. The original intent of the creators was to increase awareness of what constitutes a scientifically sound review. This approach favors guidance over sanctioning. The Canada Protocol was inspired by this type of approach. The authors tailored a checklist to their field of study, AI in mental health care and suicide prevention.

b. Design of the checklist

The authors did not identify an appropriate theoretical framework as guidance during the elaboration of the checklist. However, in order to scientifically reinforce and strengthen this study's approach, the European Society of Human Reproduction and Embryology's (ESHRE) best practices in the design of medical guidelines was used for inspiration (Vermeulen, 2018): 1. Choose a topic, 2. Create a group to develop the guidelines, 3. Scope the guidelines, 4. Formulate Key Questions, 5. Find evidence, 6. Evidence Synthesis, 7. Recommendation development, 8. Writing the guideline draft, 9. Stakeholder Consultation, 10. Approval, 11. Publication, Dissemination and Implementation.

In developing our checklist, we followed most of the cited steps. Some were excluded because the Delphi consultation already corresponded to steps 9 and 10 and as previously mentioned, this study does not create guidelines, but a checklist built on guiding ethical principles.

After having chosen the topic (step 1), the development group composed of two researchers was created: one specialized in psychology and new technologies (Carl-Maria Mörch, lead author) and one specialized in ethics and AI (Abhishek Gupta). They met on a regular basis (step 2), twice a month for six months. The initial work consisted of gathering existing recommendations from international reports on AI and ethics (n=10), important articles on this topic, and declarations in ethics in mental health and new technologies (n=3). It was determined that developing new recommendations on this topic would be less pertinent than regrouping and selecting existing recommendations and guidelines formulated by experts. This approach addresses a current problem in psychology: too many guidelines exist and it has become unclear which ones to trust or favour (Drife, 2010). By following and promoting existing recommendations, the authors hoped to increase the relevance of their work. The team read and extracted each mention of ethical issues in all documents concerning Artificial Intelligence. Overall, 450 potential ethical challenges, biases and risks were identified. After eliminating duplicate items, all the non-pertinent or redundant items were filtered. The items that were too specific, unclear, or not based on an ethical consideration (e.g. too technical) were considered non-pertinent. The remaining 329 items were divided into 9 categories: Fairness and Biases, Introductory Questions, Methodological Issues, Social Relevance and Validity, Transparency & Explainability, Controllability & Security, Autonomy, Responsibility. Of these 329 items, the lead researcher excluded all the articles that were not on Artificial Intelligence, were not applicable to the field of suicide prevention and mental health, were too technical for mental health professionals, or recommendations that were only applicable to the military (e.g. drones). 285 items were excluded and only 44 were retained. A scientific supervisor specializing in new technologies and ethics in suicide prevention (Brian Mishara) revised the checklist, its formulation and categorization. The first version of the checklist was comprised of 5 categories: "Description of the Autonomous Intelligent System" (n=12), "Privacy and Transparency" (n=8), "Security" (n=6), "Health-Related Risks" (n=9), "Biases" (n=9).

c. Theoretical context

The Canada Protocol Checklist's approach has two main theoretical frameworks: First, its goal is to facilitate responsible and civic education on new technologies, as expressed in the "Civic Media" movement from MIT's Media Lab (Zuckerman, 2014). It is interested in the use of new technologies to promote social change and enhance civic participation, including in health care.

Secondly, this study uses critical studies on BD and AI (Boyd & Crawford, 2012; Ménard, Mondoux, Ouellet, & Bonenfant, 2016) to consider technological innovations from a scientific and epistemological point of view. One of its objectives is to analyze the use of technologies rather than the discourse concerning them. For instance, the discourse concerning BD and AI often considers them to be promising and revolutionary, but their practice can also be criticized for its lack of awareness on potential biases and ethical challenges (Bowker, 2014).

d. The Delphi Methodology

In order to reinforce the internal validity of the checklist, the authors used the Delphi consultation method, developed in the 1950's by the RAND (Research and Development) Foundation (Letrilliart & Vanmeerbeek, 2011) and since used regularly in health care. It is an iterative method that consists of 2-or-more rounds of consultation with experts on a specific topic or tool. Typically, the expert panel has to study content submitted by a research team, and then give their opinion anonymously. In each round, the experts see the aggregated results, and sometimes a recall of his or her own responses. The procedure stops when a consensus has been reached (2011). A panel has been considered to require a minimum of 15 participants (Hasson, Keeney, & McKenna, 2000).

For this study, the authors determined that two rounds would suffice, because the content was already conceived and the task required was straightforward: recommending to reject, exclude, or modify (with justification). In the present study, participants were not asked to create content, but only to agree on content. The amount of rounds is usually determined by the initial needs (Ludwig, 1994); the more complex the needs, the higher the number of rounds required.

e. Experts, users and professionals consulted

To create a panel of experts:

- A list of experts (in Mental Health, Suicide Prevention, AI and New Technologies, as well as AI users) was established (see results section) who were sent individual invitations (see results section)
- Two Canadian academic research organizations were contacted that specializes in a. health care, b. health care and new technologies, and c. ethics and AI. Group invitations were sent through their mailing lists.

f. Consultation process

**First step:** An initial list of 47 items was established, divided into 5 categories: "Description of the Autonomous Intelligence System", "Privacy and Transparency", "Security", "Health-Related Risks", "Biases". The consultation was conducted using the Lime Survey online software, hosted on the Université du Québec à Montréal's servers (UQAM). The consent form was accessible on the first page of the survey, as a PDF in English. If the response to the consent form was negative, it was not possible to participate in the survey. The survey was anonymous, (Linstone & Turoff, 2002), although participants were requested to provide an email address to be contacted for the second round, which would not be associated with their responses in order to maintain anonymity.

The instructions for the first round were to: a. approve (explain the reason), b. exclude (explain the reason), or c. modify (explain the reason). The data collection phase lasted 60 days, from April 15, 2018 to June 15, 2018.

At the end of this first phase, the research team wrote a report with all the modifications completed, along with a detailed summary of the number of items excluded, modified, or kept as is. To be retained, an item had to have a score of 80% agreement or higher at the "kept as is" option (Jorm, 2015). To be excluded, an article needed to have 50% of agreement or less as "kept as is" or 80% or more to "exclude the item".

**Second step:** an edited survey was sent to the participants who provided their email address (16 participants). The summary and report were included in the invitation email. The second Delphi round requested the experts to revise the modifications and verify if the updates were satisfactory.

Participants were then asked to either keep each item or exclude it. This step followed the recommendations of the COSMIN checklist (Mokkink et al., 2010).

## Results

### a. Panel

32 experts were invited to participate by individualized emails. These participants were: a. professors, experts and researchers in suicide prevention and ICT (n=12), b. professors and researchers who have published on AI in Mental Health (n=10), c. researchers who have published on ethics and AI (n=3), d. researchers in psychology or psychiatry, experts in the use of ICT (n=2), e. experts in ethics and AI (n=2), f. professor of computer engineering specialized in AI (n=1), f. two entrepreneurs working on the use of AI in mental health and suicide Prevention (n=2).

### b. Response

16 responses were collected in the first round. This is above the suggested minimum of 15 for a first round of a Delphi survey (Hasson et al., 2000). The participants were distributed as follows : 5 were specialized in AI and mental health, 5 were specialized in technologies and mental health, 6 were specialized in suicide prevention and new technologies.

The two aforementioned Canadian research groups distributed a collective invitation to participate in their mailing lists. The group specialized in ICT, media studies and health has a mailing list of over 100 members. The group specialized in ethics and AI has a mailing list of 49 researchers across Canada.

Individual invitations have a potential response rate of 40%. In the second round, the 16 participants of the first round were invited, 8 participated. Which corresponded to a response rate of 50%.

c. Final checklist

The initial checklist included 44 items. After the first round of consultation, 12 items were retained as is or slightly modified, (condition: 80% or more of agreement at the "keep the item as is" option), 27 items were modified (condition: between 50% and 80% of agreement at the option "keep the item as is" or at the option "modify the item"), 6 items were excluded (less than 50% of agreement on "keep as is", or more than 50% at the option "remove the item"). The second round consisted of a final review of modifications of the 27 items, conducted by 8 of the 16 original participants partook in the second round. The final checklist consisted of 40 items (see Table 1):

- "Description of the Autonomous Intelligent System" (n=8)
- "Privacy and Transparency" (n=8)
- "Security" (n=6)
- "Health-Related Risks" (n=8)
- "Biases" (n=8)

d. Relevance and target users of the checklist

In the first round of the Delphi, the 16 participants were asked if they thought this instrument was relevant, using a Likert Scale ranging from 1 to 5 (1 being not relevant to 5 being very relevant). On average, the MHSP was considered relevant by the participants with an average of 4.06 out of 5 ($\sigma$=0.68).

When asked who could use this checklist, participants could not categorically assert who could use it. They were mostly uncertain for researchers (56.25%), AI developers (81.25%) or mental health professionals (68.75%). The rejection rate ("No") was two times 0% (for researchers and AI developers), and once 18.75% (for mental health professionals). This might indicate that the ethical considerations of the checklist might not be relevant to the day-to-day jobs of mental health professionals. The detailed results (see Table 2) indicate that this checklist is relevant, but might require in the future to be tailored depending on who could use this tool.

*Table 1 : Final questions*

| Question | Yes | No | Uncertain |
| --- | --- | --- | --- |
| Will researchers use this checklist ? | 7 (43.75%) | 0 (0%) | 9 (56.25%) |
| Will AI developers use this checklist ? | 3 (18.75%) | 0 (0%) | 13 (81.25%) |
| Will mental health professionals use this checklist? | 2 (12.5%) | 3 (18.75%) | 11 (68.75%) |

*Table 2 : The Canada Protocol – MHSP*

(see table file)

## Discussion

In all the studies and reviews on AI and ethics, the research team identified only three ethical tools. None of them are specialized in mental health or suicide prevention. The first, "Geneth: A General Ethical Dilemma Analyzer" (Anderson & Anderson, 2014) is a general ethical dilemma analyzer using Machine Learning. The second ethics tool is called "DELICATE: A Checklist for Trusted Learning Analytics" (Drachsler & Greller, 2016). This checklist asks the user or reader two or three questions concerning 8 key actions: determination, explain, legitimate, involve, consent, anonymize, technical, external. The third tool is a simple specialized checklist, made available on Medium website, on the policy design process including some ethical challenges: "A Canadian Algorithmic Impact Assessment" (Karlin, 2018). These tools are not directly relevant for most professionals in mental health. The Geneth tool may also be too technically complex for most users. The DELICATE checklist could be applied to a large number of situations, but was initially intended to facilitate a trusted implementation of Learning Analytics. The initial intent of this tool might make it a bit off-target for this study's requirements. For all these reasons, it seems that the Canada Protocol can serve two different current needs: 1. A tool promoting strong ethical principles when using AI or BD and providing insights on what are the ethical challenges in mental health and suicide prevention; how to identify and prevent them, 2. A tool that can be used and understood by most health care professionals and computer engineers.

Some of the core principles of the Canada Protocol are similar to the ethical guidelines of Luxton for the use of Artificial Intelligence Care Providers (AICP)(Luxton, 2014). For example, the author made recommendations for the design of an AICP. Several address similar issues as the Canada Protocol, such as " 2. Identify and provide specifications of use and limits of autonomy of AICP systems to end users" or "5. Provide built-in safeguards to assure that systems are only able to provide services within established boundaries of competence and domain of use".

## Limitations

Delphi consultations usually aim to gather expert opinions on a specific topic. This notion of expert opinions has long been discussed and debated. Some researchers consider it as a subjective perception (Linstone & Turoff, 2002). Therefore, it is acknowledged that this Delphi Consultation has taken into consideration some subjective opinions.

One way to reinforce the validity of the tool could be to conduct another Delphi consultation on the applicability of the checklist. This could include a larger committee of experts, including more computer engineering and ethics specialists.

Another limitation is that it was difficult to design a checklist that could be accessible and understood by the largest audience possible. By doing so, the authors had to make compromises. Some terms were considered too technical for some reviewers who identified as clinician or researchers in psychology. More common synonyms were used.

## Conclusion

To date, the authors have not found validated ethical guidance tools on AI in mental health or suicide prevention. Considering the sudden rise of AI in society and in health care, they developed and proposed an ethical checklist to help identify potential ethical risks, biases, and challenges: the Canada Protocol - MHSP . Checklists are commonly used in health care and it was assumed that developing a familiar tool could increase its appeal and utility. In order to validate its content, this study used a two-round Delphi Consultation. The final checklist is composed of 38 items, divided into five categories: Description of the Autonomous Intelligent System" (n=8), "Privacy and Transparency" (n=8), "Security" (n=6), "Health-Related Risks" (n=8), "Biases" (n=8).

# Références